\documentclass[aps,pra,twocolumn,superscriptaddress,longbibliography]{revtex4-1}

\usepackage{amsmath}
\usepackage{amsfonts}
\usepackage{amssymb}
\usepackage{graphicx}
\usepackage{color}
\usepackage{accents}
\usepackage[colorlinks=true, citecolor=blue]{hyperref}

\newcommand{\ket}[1]{\left|#1\right\rangle}
\newcommand{\bra}[1]{\left\langle #1\right|}

\makeatletter 
\newsavebox{\@brx}
\newcommand{\llangle}[1][]{\savebox{\@brx}{\(\m@th{#1\langle}\)}%
  \mathopen{\copy\@brx\kern-0.5\wd\@brx\usebox{\@brx}}}
\newcommand{\rrangle}[1][]{\savebox{\@brx}{\(\m@th{#1\rangle}\)}%
  \mathclose{\copy\@brx\kern-0.5\wd\@brx\usebox{\@brx}}}
\makeatother

\newlength{\dhatheight} 

\newcommand{\qed}{\nobreak \ifvmode \relax \else
      \ifdim\lastskip<1.5em \hskip-\lastskip
      \hskip1.5em plus0em minus0.5em \fi \nobreak
      \vrule height0.75em width0.5em depth0.25em\fi}

\begin{document}

\title{Simulating quantum circuits by adiabatic computation: improved spectral gap bounds} 
\author{Shane Dooley}
\email[]{dooleysh@gmail.com}
\affiliation{Dublin Institute for Advanced Studies, School of Theoretical Physics, 10 Burlington Rd, Dublin, Ireland}
\author{Graham Kells}
\email[]{gkells@stp.dias.ie}
\affiliation{Dublin Institute for Advanced Studies, School of Theoretical Physics, 10 Burlington Rd, Dublin, Ireland}
\author{Hosho Katsura}
\email[]{katsura@phys.s.u-tokyo.ac.jp}
\affiliation{Department of Physics, Graduate School of Science, The University of Tokyo, Hongo, Tokyo 113-0033, Japan}
\affiliation{Institute for Physics of Intelligence, The University of Tokyo, 7-3-1 Hongo, Tokyo 113-0033, Japan}
\author{Tony C. Dorlas}
\email[]{dorlas@stp.dias.ie}
\affiliation{Dublin Institute for Advanced Studies, School of Theoretical Physics, 10 Burlington Rd, Dublin, Ireland}
\date{\today}

\begin{abstract}  
  Adiabatic quantum computing is a framework for quantum computing that is superficially very different to the standard circuit model. However, it can be shown that the two models are computationally equivalent. The key to the proof is a mapping of a quantum circuit to an an adiabatic evolution, and then showing that the minimum spectral gap of the adiabatic Hamiltonian is at least inverse polynomial in the number of computational steps $L$. In this paper we provide two simplified proofs that the gap is inverse polynomial. Both proofs result in the same lower bound for the minimum gap, which for $L \gg 1$ is $\min_s\Delta \gtrsim \pi^2 / [8(L+1)^2]$, an improvement over previous estimates. Our first method is a direct approach based on an eigenstate ansatz, while the the second uses Weyl's theorem to leverage known exact results into a bound for the gap. Our results suggest that it may be possible to use these methods to find bounds for spectral gaps of Hamiltonians in other scenarios.
\end{abstract}


\maketitle

\section{Introduction}

Aharonov and coworkers \cite{Aha-08} proved that any quantum circuit can be efficiently simulated by an adiabatic quantum computation. Since the converse was already known \cite{Far-00}, this amounted to a proof that the circuit model and the adiabatic model are computationally equivalent. The important ingredient in the proof is Feynman's circuit-to-Hamiltonian construction, which enables the mapping of a quantum circuit to a time-independent Hamiltonian \cite{Fey-86}. This can then be used to construct an adiabatic evolution that encodes the output of the circuit in its final ground state. However, if the minimum gap during the adiabatic evolution is exponentially small in the number of computational steps $L$, it will take an exponentially long time to reach the final ground state. Hence, to show that the the circuit is \emph{efficiently} simulated by the adiabatic evolution it is also necessary to show that the minimum spectral gap is at least inverse polynomial in $L$. In Ref. \cite{Aha-08} this was achieved by deriving a lower bound for the minimum gap, $\min_s \Delta(s) \geq 1/144 L^2$. However, the derivation is quite complicated, and involves using Gerschgorin's Circle Theorem and a conductance bound from the theory of rapidly mixing Markov chains. In subsequent work, Deift, Ruskai and Spitzer \cite{Dei-07} gave an improved bound $\min_s\Delta(s) > 1/[2(L+1)^2]$. An alternative proof of the computational equivalence that does not rely on Feynman's circuit-to-Hamiltonian construction was given in Ref. \cite{Miz-07}.

In this paper we provide two relatively simple proofs that the minimum gap is bounded by: \begin{eqnarray} \min_s \Delta(s) &\geq& 2\sqrt{2\left[1 + \cos\epsilon \right]} - 2 \left[1 + \cos\epsilon\right] \nonumber \\ &=& \frac{\epsilon^2}{2} + \mathcal{O}(\epsilon^{4}) , \nonumber \end{eqnarray} where $\epsilon = \pi / (2L+2)$. Our first proof of the bound is based on an ansatz for the eigenstates of the adiabatic Hamiltonian, from which we find good approximations to the eigenstates and to the full spectrum of eigenvalues. The second proof involves a decomposition of the adiabatic Hamiltonian into an appropriate sum of Hermitian operators, followed by an application of Weyl's theorem, which gives bounds on the eigenvalues of the adiabatic Hamiltonian. The two different methods of derivation are seemingly unrelated, but surprisingly give the same lower bound for the spectral gap.

The paper is outlined as follows. In section \ref{sec:FeynmansH} we provide some background to the circuit model and Feynman's mapping of the circuit to a Hamiltonian evolution. In section \ref{sec:AQC} we give a brief review of adiabatic quantum computing, and the problem of simulating a circuit with an adiabatic evolution. Then, in section \ref{sec:first_proof} we give our first derivation of the bound above and in section \ref{sec:second_proof} we give our second derivation of the bound.


\section{Feynman's clock Hamiltonian}\label{sec:FeynmansH}

In the circuit model of quantum computing, a calculation is implemented in several stages. First, a set of $N$ logical qubits are prepared in the computational basis state $\ket{\alpha(0)} = \ket{0_10_2\cdots 0_N}$. Next, a sequence of one- or two-qubit gates are applied so that after a total of $L$ gates $U_1 ,..., U_L$ the system is in the output state $\ket{\alpha(L)} = U_L\cdots U_1 \ket{\alpha(0)}$ (intermediate states are denoted $\ket{\alpha(l)} = U_l \cdots U_1 \ket{\alpha(0}$, with $l=1,...,L$). Finally, the output state is measured in the computational basis \cite{Deu-89, Nie-00}.

Although the computation is implemented by a discrete sequence of unitaries, Feynman showed that it is possible to map the circuit to a continuous time evolution with a time-independent Hamiltonian \cite{Fey-86}. This can be done by adding to the $N$ logical qubits an $L+1$ dimensional ancillary ``clock'' system, and constructing the Hamiltonian: \begin{eqnarray} \mathsf{H}_c = \frac{1}{2}\sum_{l=1}^{L} &\bigg(& I \otimes \ket{l-1}_c\bra{l-1}_c + I \otimes \ket{l}_c\bra{l}_c \nonumber\\ && - U_l \otimes \ket{l}_c\bra{l-1}_c - U_l^\dagger \otimes \ket{l-1}_c\bra{l}_c  \bigg) , \label{eq:H_c} \end{eqnarray} where $\{ \ket{l}_c \}_{l=0}^L$ are a set of basis states for the clock system. The clock is prepared in the state $\ket{0}_c$ and the system is allowed to evolve by $\mathsf{H}_c$ for a period of time. If a final measurement of the clock system in the basis $\{ \ket{l}_c \}_{l=0}^L$ gives the outcome corresponding to the state $\ket{L}_c$, then the $N$ logical qubits will be in the output state $\ket{\alpha(L)}$. We note that the first two terms in Eq. \ref{eq:H_c} are not necessary but are included for later convenience, since they ensure that $\mathsf{H}_c$ is positive semidefinite.

The Feynman circuit-to-Hamiltonian construction was used by Kitaev as part of his proof that the local Hamiltonian probelem is QMA-complete \cite{Kit-02}, and, inspired by this, it was used by Aharonov and coworkers to prove that adiabatic quantum computation is computationally equivalent to the circuit model \cite{Aha-08}. This latter application is the focus of this paper. In the next section we give a brief review of adiabatic quantum computing and a summary of the proof in Ref. \cite{Aha-08}. 

\section{Adiabatic quantum computation}\label{sec:AQC}

Adiabatic quantum computing is a framework for quantum computing that is based on adiabatic variation of a Hamiltonian $\mathsf{H}(s)$, where $s \in [0,1]$ is a tunable parameter \cite{Alb-18}. At some initial time the system is prepared in the ground state of the initial Hamiltonian $\mathsf{H}(0) = \mathsf{H}_\text{init}$. By the adiabatic theorem, if $s$ is increased slowly relative to the size of the spectral gap of $\mathsf{H}(s)$ the system will remain in the ground state of $\mathsf{H}(s)$ at each instant until it reaches the ground state of the final Hamiltonian $\mathsf{H}(1) = \mathsf{H}_\text{final}$. For example, we may consider the Hamiltonian $\mathsf{H}(s) = (1-s) \mathsf{H}_\text{init} + s \mathsf{H}_\text{final}$ where $s$ is a rescaled time parameter $s=t/T$ that increases from $s=0$ to $s=1$ as time evolves from the initial time $t=0$ to the final time $t=T$. If the output of the quantum computation can be encoded in the final ground state, an adiabatic quantum computation will have been implemented.

The relationship between the computational power of the adiabatic model and the computational power of the circuit model is not obvious. However, it was proved in Ref. \cite{Aha-08} that they are computationally equivalent. This was done by showing that any quantum computation with a circuit description can be efficiently simulated by an adiabatic quantum evolution. Since the converse was already known to be true \cite{Far-00}, this proved the computational equivalence of the two models. We now summarise the argument of Ref. \cite{Aha-08}.

Since the input to the quantum circuit is the $N$-qubit state $\ket{\alpha(0)}$ and the output is $\ket{\alpha(L)}$, it is natural to try to construct an $N$-qubit adiabatic model that has $\ket{\alpha(0)}$ as the ground state of its initial Hamiltonian $\mathsf{H}_\text{init}$ and $\ket{\alpha(L)}$ as the ground state of its final Hamiltonian $\mathsf{H}_\text{final}$. However, this approach runs into the problem that the output state $\ket{\alpha(L)}$ is unknown: how then can the corresponding Hamiltonian $\mathsf{H}_\text{final}$ be constructed? In Ref. \cite{Aha-08} it was shown that, following Feynman and Kitaev, this difficulty can be overcome by adding to the $N$ logical qubits an $L+1$ dimensional clock system. It is straightforward to verify that the choice of initial Hamiltonian $\mathsf{H}_\text{init} = \sum_{i=1}^N \ket{1_i}\bra{1_i} \otimes \ket{0}_c\bra{0}_c + I \otimes \sum_{l=1}^L \ket{l}_c\bra{l}_c$ has the desired ground state $\ket{\alpha(0)} \otimes | 0 \rangle_c$. (Here, $\ket{1_i}\bra{1_i}$ represents a projector onto the state $|1\rangle$ of the $i$'th qubit, with the identity operator acting on all other qubits.) It can also be checked that the final Hamiltonian $\mathsf{H}_\text{final} = \mathsf{H}_c + \sum_{i=1}^N \ket{1_i}\bra{1_i} \otimes \ket{0}_c\bra{0}_c$, where $H_c$ is given in Eq. \ref{eq:H_c}, has the ground state: \begin{equation} \ket{\eta} \equiv \frac{1}{\sqrt{L+1}} \sum_{l=0}^L \ket{\gamma(l)}, \nonumber \end{equation} which is a superposition of the $L+1$ orthonormal states $\ket{\gamma(l)} \equiv \ket{\alpha(l)} \otimes | l \rangle_c$. If a final measurement of the clock system in the basis $\{ \ket{l}_c \}_{l=0}^L$ gives the outcome corresponding to the state $\ket{L}_c$, then we know that the $N$ logical qubits will be in the output state $\ket{\alpha(L)}$. We emphasise that the Hamiltonian $\mathsf{H}_\text{final}$ defined above can be constructed using only knowledge of the quantum circuit $\{ U_1,...,U_L \}$, and without direct knowledge of the output state $\ket{\alpha(L)}$.



The $L+1$ dimensional subspace spanned by the states $\{ \ket{\gamma(l)} \}_{l \in \{0,...,L\}}$ is invariant under evolution by $\mathsf{H}(s)$ \cite{Aha-08}. This means that, although the full state space is $(L+1)2^{N}$ dimensional, the evolution by $\mathsf{H}(s)$ takes place entirely within the smaller $L+1$ dimensional subspace. The Hamiltonian $\mathsf{H}(s)$ restricted to this subspace is denoted $H(s)$ and is given in the $\{ \ket{\gamma(l)} \}$ basis by the $L+1$ dimensional tridiagonal matrix \cite{Aha-08}:





\begin{equation}
 H(s) =  \left( \begin{array}{cccccc}
      \frac{s}{2}  & -\frac{s}{2} & & & & \\
      -\frac{s}{2} & 1 & -\frac{s}{2} & & & \\
       & -\frac{s}{2} & 1 & & & \\
       & & & \ddots & & \\
       & & & & 1 & -\frac{s}{2} \\
       & & & & -\frac{s}{2} & 1-\frac{s}{2} \end{array} \right) . \nonumber
 \end{equation} Its eigenvalue equation is $H(s)\psi_l(s) = \lambda_l (s) \psi_l(s)$ where $l = 0,1,...,L$ and we label the eigenvalues in increasing order $\lambda_0(s) \leq \lambda_1 (s) \leq ... \leq \lambda_L(s)$. To show that the adiabatic evolution is an  \emph{efficient} simulation of the quantum circuit it must be demonstrated that the spectral gap $\Delta (s) = \lambda_1(s) - \lambda_0(s)$ is at least inverse polynomial in the number of computation steps $L$. In the following, we give two proofs that this is the case.

\section{First proof}\label{sec:first_proof}

Our first proof is based on an ansatz for the eigenvectors of $H(s)$. The ansatz leads to a set of trancendental equations. Although these equations cannot be solved explicitly, they can be used to find approximate expressions for the eigenvalues and eigenvectors, and a lower bound for the spectral gap.
 
\subsection{Ansatz}\label{sec:ansatz}

We propose an ansatz $\psi$ for the eigenvector with the vector elements $[\psi]_k = \alpha z^{k+1} + \beta z^{-k-1}$, $k=0,...,L$. For now, $\alpha$, $\beta$ and $z$ are arbitrary complex numbers, but they will be specified shortly by the requirement that $\psi$ be an eigenstate of $H$. Multiplying $\psi$ by the Hamiltonian $H$ gives a vector $H\psi$ with the elements: \begin{eqnarray} \left[ H\psi \right]_0 &=& \frac{s}{2}(\alpha z + \beta z^{-1}) - \frac{s}{2}(\alpha z^2 + \beta z^{-2}) , \nonumber \\ \left[ H \psi \right]_k  &=& \lambda [\psi]_k , \quad \lambda = 1 - \frac{s}{2}(z + z^{-1}), \quad 0<k<L, \nonumber \\ \left[ H\psi \right]_{L} &=& -\frac{s}{2}(\alpha z^{L} + \beta z^{-L}) + \left(1 - \frac{s}{2}\right)(\alpha z^{L+1} + \beta z^{-L-1}) . \nonumber \end{eqnarray} We see that $\psi$ is ``almost'' an eigenvector of $H$ with eigenvalue $\lambda = 1 - \frac{s}{2}(z + z^{-1})$, but not quite, since the first and last elements $[H\psi]_0$ and $[H\psi]_L$ do not have the correct form. Enforcing $[H\psi]_0 = \lambda [\psi]_0$ and $[H\psi]_L = \lambda [\psi]_L$ leads to the conditions $s(\alpha + \beta) = ( 2 - s ) (\alpha z + \beta z^{-1})$ and $\beta = \alpha z^{2L+3}$, respectively. Substituting the second equation into the first gives a new condition: \begin{eqnarray} s &=& \frac{2(z^{-L-1/2} + z^{L+1/2})}{z^{L+3/2} + z^{L+1/2} + z^{-L-1/2} + z^{-L-3/2}} \nonumber\\ &\equiv& f(z) , \label{eq:z_condition} \end{eqnarray} where, for later convenience, we have introduced the function $f(z)$. If $z$ is a solution to Eq. \ref{eq:z_condition}, our ansatz $\psi$ is an eigenstate of $H(s)$. We also note that the eigenvalue $\lambda = 1 - \frac{s}{2}(z + z^{-1})$ must be real, since the Hamiltonian $H$ is symmetric. This implies that there are two possibilities for $z$: either $z$ is real or $z$ is complex with unit modulus.

\begin{figure*}
  \includegraphics[width=0.9\textwidth]{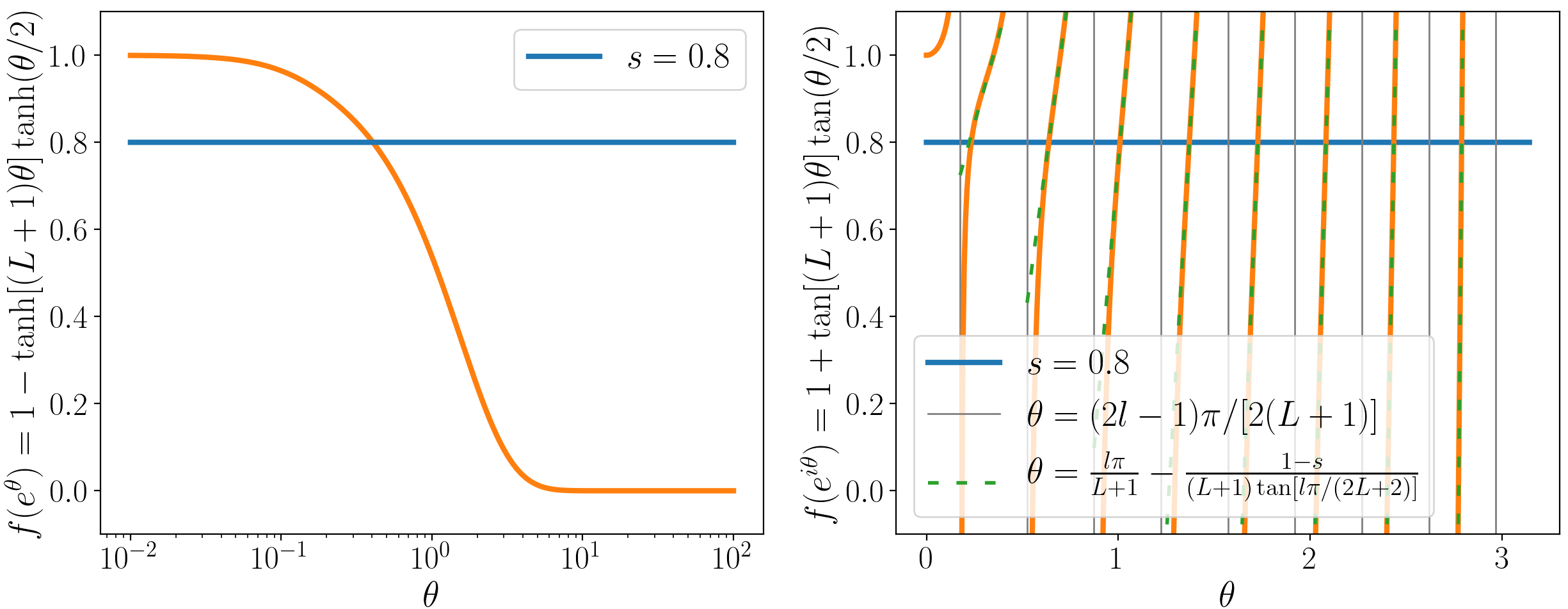}
\caption{\label{fig:f} Solutions to the trancendental equations Eq. \ref{eq:f_z_real} (left figure) and Eq. \ref{eq:f_z_complex} (right figure) are found when $f(z) = s$. Plotted for $L=8$.}
\end{figure*}

\subsubsection{Real $z$ solution}

We first consider the case where $z$ is real. Writing $z = e^\theta$ the eigenvalue and (unnormalised) eigenvector elements are: \begin{equation} \lambda = 1 - s \cosh \theta  , \quad [\psi]_k = \cosh [(L - k + 1/2)\theta] , \nonumber \end{equation} where, substituting $z=e^\theta$ into Eq. \ref{eq:z_condition}, we see that $\theta$ is a solution to the equation: \begin{equation} s = f(e^{\theta}) = 1 - \tanh [(L+1)\theta] \tanh (\theta/2) . \label{eq:f_z_real} \end{equation} Fig. \ref{fig:f}(a) shows $f(e^{\theta})$ plotted as a function of $\theta$. The plot shows (and it is easily verified by a calculation) that $f(e^\theta)$ is continuous and monotonically decreasing, and is guaranteed to intersect a horizontal line at $f(e^\theta) = s$ (at finite $\theta$ if $s>0$, and at $\theta\to\infty$ if $s=0$). This gives exactly one solution to the eigenvalue equation, which we denote $\theta_0$. The corresponding eigenvalue and eigenvector are similarly labelled $\lambda_0$ and $\psi_0$.

\subsubsection{Complex $z$ solutions}\label{sec:ansatz_z_complex}

Next, in the case where $z$ is complex with unit modulus we can write $z=e^{i\theta}$, which gives the eigenvalue and (unnormalised) eigenvector: \begin{equation} \lambda = 1 - s \cos \theta  , \quad [\psi]_k = \cos [(L - k + 1/2)\theta] , \label{eq:eig_z_complex} \end{equation} where $\theta$ is the solution to the equation: \begin{equation} s = f(e^{i\theta}) = 1 + \tan [(L+1)\theta] \tan (\theta/2) . \label{eq:f_z_complex} \end{equation} If $\theta$ is a solution to Eq. \ref{eq:f_z_complex} then it is clear that $\theta + 2 \pi m$ is also a solution for any $m\in\mathbb{Z}$, since $z = e^{i(\theta + 2 \pi m)} = e^{i\theta}$. However, replacing $\theta \to \theta + 2 \pi m$ in the expressions for the eigenvalue and eigenstate in Eq. \ref{eq:eig_z_complex} shows that these solutions all correspond to the same eigenvalue and eigenvector. We may therefore restrict to solutions to Eq. \ref{eq:f_z_complex} in any $2\pi$ range, say $\theta \in [-\pi, \pi)$. Also, if $\theta$ is a solution to Eq. \ref{eq:f_z_complex} then so is $-\theta$, since $f(e^{i\theta}) = f(e^{-i\theta})$. But again, replacing $\theta\to -\theta$ in Eq. \ref{eq:eig_z_complex} shows that this does not give a distinct solution to the eigenvalue equation. All distinct solutions may therefore be found in the range $\theta \in [0,\pi]$. In Fig. \ref{fig:f}(b) we plot $f(e^{i\theta})$ as a function of $\theta$ in this range (the solid orange lines). The function $f(e^{i\theta})$ diverges for certain values of $\theta$. From Eq. \ref{eq:f_z_complex} we see that these divergences occur at the points $\theta = (2l-1)\pi/[2(L+1)]$, for $l = 1,...,L$, marked by grey vertical lines in Fig. \ref{fig:f}(b). Moreover, in the interval $\theta \in \left[ \frac{(2l-1)\pi}{2(L+1)}, \frac{(2l+1)\pi}{2(L+1)} \right]$ between two consecutive divergences the function $f(e^{i\theta})$ increases continuously from $-\infty$ to $+\infty$ and will therefore cross a horizontal line at $f(e^{i\theta})=s$. Since there are $L$ such intervals in the range $\theta \in [0, \pi]$ we are guaranteed $L$ solutions, which we denote $\theta_l$ where $l = 1,...,L$. The corresponding eigenvalues and eigenvectors are $\lambda_l$ and $\psi_l$. Since the eigenvalue $\lambda = 1 - s\cos\theta$ is an increasing function of $\theta$ for $\theta \in [0,\pi]$ we are also guaranteed that the eigenvalues are labelled in increasing order $\lambda_1(s) \leq \lambda_2(s) \leq ... \leq \lambda_L(s)$.

Combining the solution $\theta_0$ obtained for real $z$ with the $L$ solutions $\{\theta_1, ... , \theta_L\}$ for complex $z$ gives a complete set of $L+1$ eigenvalues and eigenvectors of $H(s)$. 

\subsection{Approximations}\label{sec:approx}

Although we have identified the complete set of $L+1$ solutions to the eigenvalue equation, we cannot solve the trancendental equations \ref{eq:f_z_real} and \ref{eq:f_z_complex} to find explicit solutions. However, progress can be made by finding approximate solutions.

\subsubsection{Ground state approximation}\label{sec:ground_approx}

\begin{figure*}
  \includegraphics[width=0.49\textwidth]{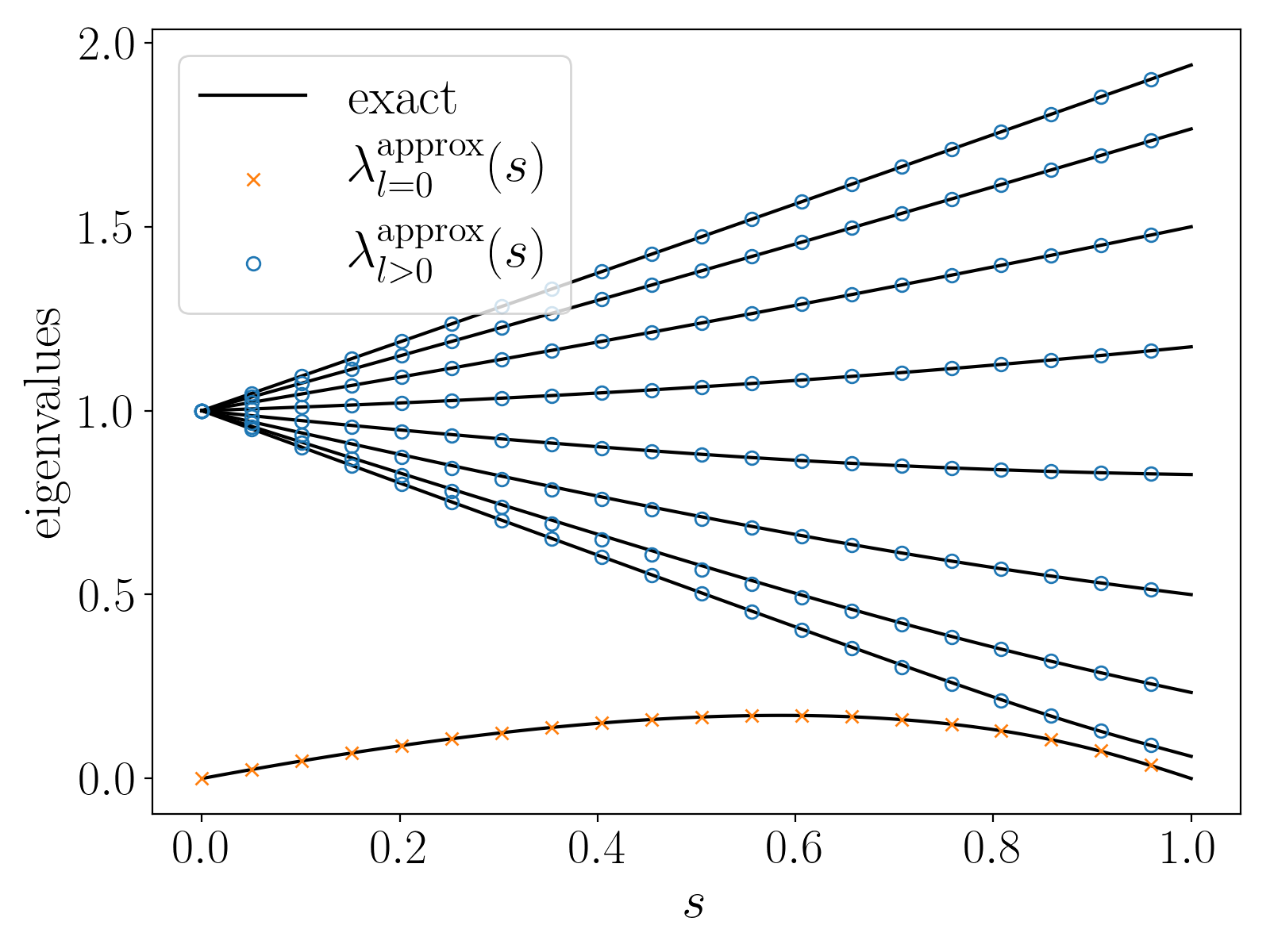}
  \includegraphics[width=0.49\textwidth]{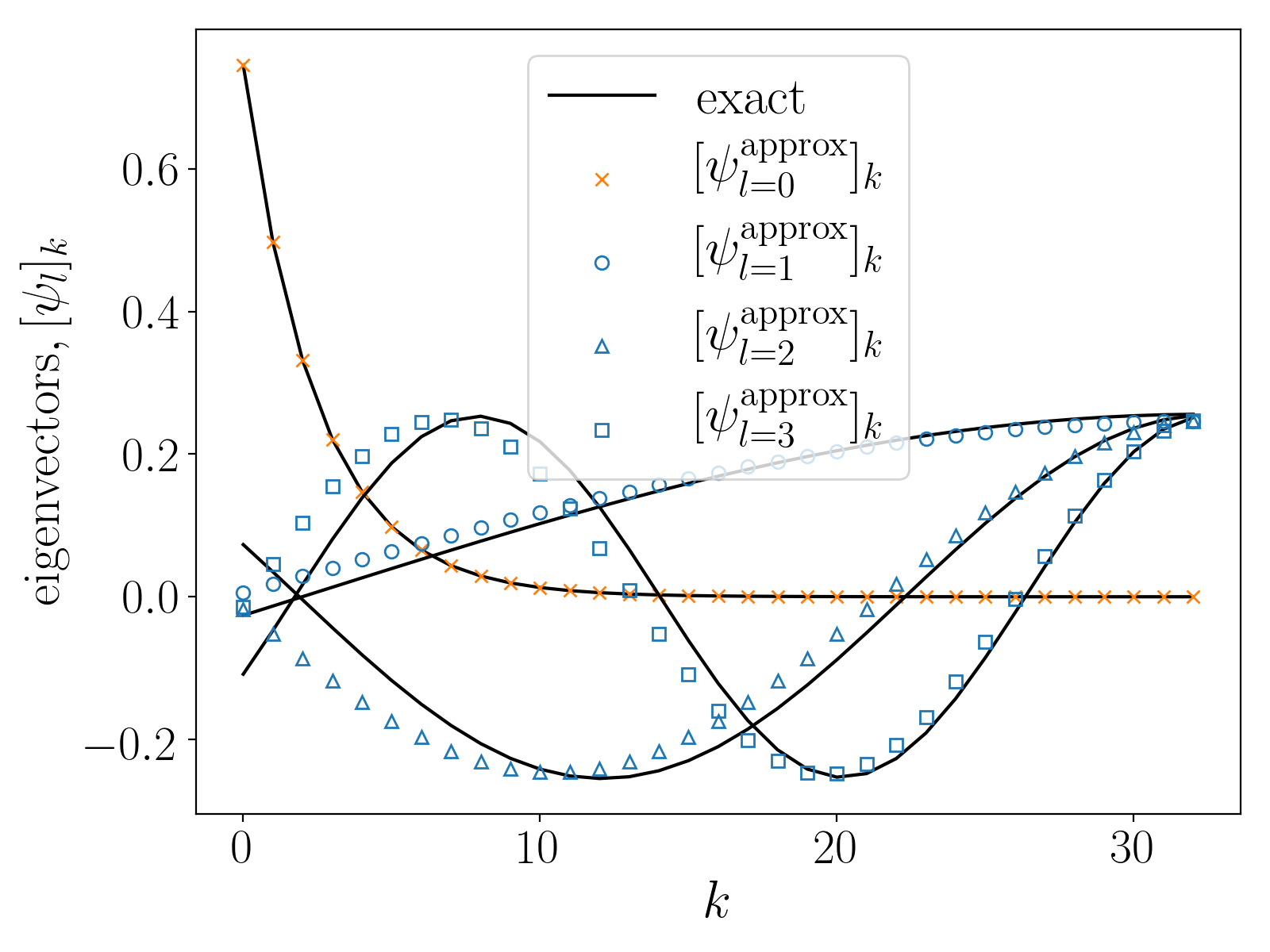}
\caption{\label{fig:exact_vs_approx_eigvals} Left: A comparison of the exact and the approximate eigenvalues for $L=8$. Right: exact vs. approximate eigenstates (the 4 lowest) for $s=0.8$ and $L=32$.}
\end{figure*}

We begin with the ground state eigenvalue $\lambda_0 = 1 - s\cosh\theta_0$ and eigenvector $[\psi_0]_k = \cosh[(L-k+1/2)\theta_0]$, where $\theta_0$ is the solution to $s = f(e^{\theta_0}) = 1 - \tanh[(L+1)\theta_0]\tanh(\theta_0 /2)$. When $(L+1)\theta_0 \gg 1$ we have $\tanh[(L+1)\theta_0] \approx 1$, which gives $\theta_0 \approx 2\tanh^{-1}(1-s) = \ln\left( \frac{2}{s}-1 \right)$. With this approximation the ground state eigenvalue is: \begin{eqnarray} \lambda_0(s) &\approx& 1 - s \cosh\left[ \ln\left( \frac{2}{s}-1 \right) \right] \nonumber \\ &=& \frac{s(1-s)}{2 - s} \equiv \lambda_0^\text{approx}(s) , \label{eq:gs_ub} \end{eqnarray} and the eigenvector elements are: \begin{eqnarray} [\psi_0]_k &\approx& \frac{1}{2} \left[ \left( \frac{2}{s} - 1 \right)^{L-k+1/2} + \left( \frac{2}{s} - 1 \right)^{-L+k-1/2} \right] \nonumber \\ &\equiv& [\psi_{0}^{\text{approx}}]_k .\nonumber \end{eqnarray} In Fig. \ref{fig:exact_vs_approx_eigvals} we compare the exact ground state and its eigenvalue (found by numerical diagonalisation of $H$) with the approximation.

We note that the approximation $\tanh[(L+1)\theta_0] \approx 1$ overestimates $\tanh[(L+1)\theta_0]$, and so underestimates $\theta_0$. The approximation therefore overestimates the eigenvalue $\lambda_0(s)$ and is an upper bound to the true eigenvalue, $\lambda_0^\text{approx}(s) \geq \lambda_0(s)$.

\subsubsection{Excited state approximations}\label{sec:excited_approx}

We next approximate the excited state eigenvalues $\lambda_l = 1 - s\cos\theta_l$ and eigenvectors $[\psi_l]_k = \cos[(L-k+1/2)\theta_l]$, where $\theta_l$ are the $L$ solutions to $s = f(e^{i\theta}) = 1 + \tan[(L+1)\theta]\tan(\theta /2)$ in the range $\theta \in (0, \pi)$. In section \ref{sec:ansatz} we showed that the $l$'th solution $\theta_l$ lies in the range $\theta_l \in \left[ \frac{(2l-1)\pi}{2(L+1)}, \frac{(2l+1)\pi}{2(L+1)} \right]$ between two consecutive divergences of $f(e^{i\theta})$. A lower bound to the true value of $\theta_l$ is therefore found by choosing the smallest value in this range, i.e., $\theta_{l} \approx (2l-1)\pi / [2(L+1)]$. Since $\lambda = 1 - s \cos\theta$ is an increasing function of $\theta$ for $\theta \in [0,\pi]$, a lower bound to the eigenvalue $\lambda_l$ can be obtained using this lower bound for $\theta_l$: \begin{equation} \lambda_l(s) \geq 1 - s \cos\frac{(2l-1)\pi}{2(L+1)} . \label{eq:es_lb} \end{equation} This lower bound will be useful in the next subsection when we derive a lower bound on the spectral gap. However, in terms of the error with respect to the true value of $\theta_l$, the approximation $\theta_l \approx (2l-1)\pi / [2(L+1)]$ begins to break down as $s$ increases or as $l$ increases. This can be seen in Fig. \ref{fig:f}(b), where the approximations (the thin vertical grey lines) depart from the true values (the solid orange lines) as $s$ increases or as $l$ increases. However, the approximation may be improved by observing that for $s \approx 1$ or for $l \approx L$, the function $f(e^{i\theta})$ is well approximated by the linear expansion around the solutions to $f(e^{i\theta}) = 1$, i.e., around $\theta = l\pi/(L+1)$. This gives: \begin{equation} s = f(e^{i\theta}) \approx 1 + \left[ \theta - \frac{l\pi}{L+1} \right] (L+1) \tan\frac{l\pi}{2(L+1)} , \nonumber \end{equation} which we can easily solve for $\theta$ to obtain $\theta_l \approx l\pi/(L+1) - (1-s)/[(L+1)\tan[l\pi/2(L+1)]$. These linear approximations are plotted in the green dashed lines in Fig. \ref{fig:f}(b). The two approximations for $\theta_l$ may be combined in a single approximation: \begin{eqnarray} \theta_l &\approx& \max\left\{ \frac{(2l-1)\pi}{2(L+1)}, \: \frac{l\pi}{L+1} - \frac{1-s}{(L+1)\tan[l\pi/(2L+2)]} \right\} \nonumber \\ &\equiv& \theta_{l}^{\text{approx}} . \nonumber \end{eqnarray} In Fig. \ref{fig:exact_vs_approx_eigvals} we compare the approximate eigenvectors and eigenvalues with the exact ones (found numerically).

\subsubsection{Lower bound for the gap}

In Eq. \ref{eq:gs_ub} we found that an upper bound to the ground state eigenvalue is $\lambda_0(s) \geq s(1-s)/(2-s)$, and in Eq. \ref{eq:es_lb} that a lower bound to the first excited state eigenvalue is $\lambda_1 (s) \leq 1 - s\cos\epsilon$, where $\epsilon = \pi/(2L+2)$. This means that the spectral gap is bounded by: \begin{equation} \Delta (s) \geq 1 - s \cos\epsilon - \frac{s(1-s)}{2-s} . \end{equation} Minimising over $s$ gives: \begin{equation} \min_s \Delta (s) \geq 2\sqrt{2 \left[ 1 + \cos\epsilon \right]} - 2 \left[ 1 + \cos\epsilon \right] = \frac{\epsilon^2}{2} + \mathcal{O}(\epsilon^4) , \label{eq:gap_bound} \end{equation} where we have expanded the right hand side to leading order in $\epsilon$. This proves that the spectral gap of $H(s)$ is at least inverse quadratic in $L$.

We now give a second derivation of the same bound, based on an application of Weyl's theorem, leaving a discussion of the result to the conclusion in section \ref{sec:conclusion}.

\section{Second proof}\label{sec:second_proof}

Our second proof is based on an application of Weyl's theorem to an appropriate decomposition of the Hamiltonian $H(s)$.


\subsection{Weyl's theorem}

Weyl's theorem is an inequality on the eigenvalues of Hermitian matrices and their sums \cite{Fra-12}. Let $A$ and $B$ be $(L+1) \times (L+1)$ Hermitian matrices, and denote by $\{ \mu_j (A) \}_{j=0}^L$, $\{ \mu_j (B) \}_{j=0}^L$, and $\{ \mu_j (A+B) \}_{j=0}^L$ the sets of eigenvalues of $A$, $B$, and $A+B$, respectively, with the eigenvalues labelled in increasing order. Weyl's theorem says that for any $0 \leq k \leq L$: \begin{equation} \mu_k(A) + \mu_0(B) \leq \mu_k(A+B) \leq \mu_k(A) + \mu_L(B) . \label{eq:Weyl} \end{equation}

Moreover, if $B$ is positive semidefinite we have $\mu_k(A) \leq \mu_k(A) + \mu_0(B)$. Therefore, a corollary of Weyl's theorem that if $B$ is positive semidefinite, we have $\mu_k(A) \leq \mu_k (A+B)$ for all $0 \leq k \leq L$. Alternatively, making the substitution $A' = A+B$, we have: \begin{equation} \mu_k(A' - B) \leq \mu_k (A') , \label{eq:Weyl_corollory} \end{equation} for all $0 \leq k \leq L$ if $B$ is positive semidefinite.

Weyl's theorem suggests a strategy for deriving a bound on the spectral gap of $H(s)$: we should find a decomposition $H(s) = A + B$ for which the eigenvalues of $A$ and $B$ are known. Then, by applying the theorem we can find an upper bound to the ground state and a lower bound to the first excited state of $H(s)$. We begin by introducing the necessary analytically tractable matrices that will play the roles of $A$ or $B$.

\subsection{Analytically tractable tridiagonal matrices}

We define the $(L+1) \times (L+1)$ real matrix: \begin{equation}
 T(a,b) =  \left( \begin{array}{cccccc}
      a  & 1 & & & & \\
      1 & 0 & 1 & & & \\
       & 1 & 0 & & & \\
       & & & \ddots & & \\
       & & & & 0 & 1 \\
       & & & & 1 & b \end{array} \right) . \nonumber
 \end{equation} For certain special choices of the numbers $a$ and $b$ it is possible to calculate the eigenvalues of this matrix, using the eigenstate ansatz given at the beginning of section \ref{sec:ansatz}. For our purposes, we need only two cases. First, the matrix $T(-1,1)$ (with $a=-1$ and $b=1$), has the eigenvalues \cite{Yue-05, Wil-08, Koa-06}: \begin{equation} \mu_j (T(-1,1)) = 2\cos [(2j+1)\epsilon] , \qquad 0 \leq j \leq L , \label{eq:ev1} \end{equation} where $\epsilon = \pi/(2L+2)$ Second, the matrix $T(q,q^{-1})$ (with $a=q$ and $b = q^{-1}$) has the eigenvalues \cite{Sal-90, Wil-08, Kat-15}: \begin{eqnarray} \mu_0(T(q,q^{-1})) &=& q + q^{-1} , \label{eq:ev0} \\ \mu_j(T(q, q^{-1})) &=& 2\cos (2 j \epsilon), \qquad 1 \leq j \leq L . \end{eqnarray}
 
 \subsection{Lower bound for the gap}

 To find a lower bound to the first excited state eigenvalue, we first write: \begin{equation} H(s) = \mathbb{I}_{L+1} - \frac{s}{2}T(-1, 1) - \ket{\gamma(0)}\bra{\gamma(0)} . \end{equation} This decomposition is of the form $H(s) = A + B$ where $A = - \ket{\gamma(0)}\bra{\gamma(0)}$ and $B = \mathbb{I}_{L+1} - \frac{s}{2}T(-1, 1)$ are Hermitian. Thus, we can apply Weyl's theorem (Eq. \ref{eq:Weyl}) to write the inequality: \begin{equation} \mu_1(- \ket{\gamma(0)}\bra{\gamma(0)}) + \mu_0(\mathbb{I}_{L+1} - \frac{s}{2}T(-1, 1)) \leq \mu_1(H(s)) . \end{equation} Now, we can use Eq. \ref{eq:ev1} to find $\mu_0(\mathbb{I}_{L+1} - \frac{s}{2}T(-1, 1)) = 1 - s \cos\epsilon$. Since $\mu_1(- \ket{\gamma(0)}\bra{\gamma(0)}) = 0$, we then have: \begin{equation} \lambda_1(s) \geq 1 - s \cos\epsilon , \end{equation} where we have reverted to the notation $\lambda_1(s) = \mu_1 (H(s))$ for eigenvalues of $H(s)$. Note that this is the same as the lower bound derived in Eq. \ref{eq:es_lb}.
 
 Next, we find an upper bound to the ground state eigenvalue of $H(s)$. We write: \begin{equation} H(s) = \mathbb{I}_{L+1} - \frac{s}{2}T(q, q^{-1}) - \frac{s(s-1)}{2-s}\ket{\gamma(L)}\bra{\gamma(L)} , \nonumber \end{equation} where $q = \frac{2}{s} - 1$. This decomposition is of the form $H(s) = A' - B$ where $A' = \mathbb{I}_{L+1} - \frac{s}{2}T(q, q^{-1})$ is Hermitian and $B =  \frac{s(s-1)}{2-s}\ket{\gamma(L)}\bra{\gamma(L)}$ is positive semidefinite. Thus, we can apply the corollory to Weyl's theorem (Eq. \ref{eq:Weyl_corollory}) to write: \begin{equation} \mu_0 (H(s)) \leq \mu_0 (\mathbb{I}_{L+1} - \frac{s}{2}T(q, q^{-1})) . \end{equation} This gives \begin{equation} \lambda_0(s) \leq \frac{s(1-s)}{2-s} , \end{equation} where we have used Eq. \ref{eq:ev0} to determine the right hand side of the inequality. Note that this is the same as the upper bound derived in Eq. \ref{eq:gs_ub}.

 Since the bounds are the same as those derived in section \ref{sec:first_proof}, they combine to give the same lower bound for the minimum spectral gap, given in Eq. \ref{eq:gap_bound}. \vspace{4mm}

 \section{Conclusion}\label{sec:conclusion}

 For the simulation of a quantum circuit with $L$ computation steps by an adiabatic evolution by a Hamiltonian $\mathsf{H}(s)$, we have presented two relatively simple proofs that the spectral gap of $\mathsf{H}(s)$ is at least inverse quadratic in $L$. Both proofs result in identical lower bounds for the minumum spectral gap: \begin{equation} \min_s \Delta (s) \geq 2\sqrt{2 \left[ 1 + \cos\epsilon \right]} - 2 \left[ 1 + \cos\epsilon \right] = \frac{\epsilon^2}{2} + \mathcal{O}(\epsilon^4) , \nonumber \end{equation} where $\epsilon = \pi/(2L+2)$. In the large $L$ limit, our lower bound is a factor of $18\pi^2 \sim 177$ times larger than the bound $\min_s \Delta(s) \geq 1/144L^2$ that is given in Ref. \cite{Aha-08}. Moreover, both proofs here are more elementary than the one presented in Ref. \cite{Aha-08}. With respect to the lower bound $\min_s \Delta(s) \geq 1/[2(L+1)]^2$ given in Ref. \cite{Dei-07}, ours is an improvement by a factor of $\pi^2 / 4 \sim 2.5$ in the large $L$ limit.

 Our first proof (section \ref{sec:first_proof}) is based on an eigenstate ansatz that leads to a set of trancendental equations, reminiscent of the Bethe ansatz. Our second proof (section \ref{sec:second_proof}) is based on an application of Weyl's theorem. Both approaches open up the possibility of using these tools in deriving bounds on spectral gaps in generalisations of the adiabatic Hamiltonian considered here.


\begin{acknowledgments}
We would like to thank Ian Jubb, Luuk Coopmans and Kevin Kavanagh for helpful discussions. S.D. and G.K. acknowledge support from Science Foundation Ireland through Career Development Award 15/CDA/3240. H.K. was supported in part by JSPS KAKENHI Grants No. JP18H04478 and No. JP18K03445.
\end{acknowledgments}


\bibliography{refs}

\end{document}